\documentclass[%
reprint, 
amsmath,amssymb, 
aps,
]{revtex4-2} 

\usepackage{graphicx}
\usepackage{dcolumn}
\usepackage{bm}
\usepackage[dvipsnames]{xcolor}

\begin{document}
\preprint{APS/123-QED}

\title{Strongly interacting 2D electron systems: Evidence for enhanced 1D edge-channel coupling}

\author{C. Marty$^{1,2}$, C. Reichl$^{1,2}$, S. Parolo$^{1,2}$, I. Grandt-Ionita$^{1,2,3}$, J. Scharnetzky$^{1,2}$, W. Dietsche$^{1,2,4}$ and W. Wegscheider$^{1,2}$}
\affiliation{$^1$Laboratory for Solid State Physics, ETH Zürich, Zürich, Switzerland\\$^2$Quantum Center, ETH Zürich, Zürich, Switzerland\\$^3$Institut für Nanostruktur- und Festkörperphysik, Universität Hamburg, Hamburg, Germany\\
$^4$Max-Plank-Institut für Festkörperforschung, Stuttgart, Germany}

\begin{abstract}
    We observe nearly vanishing Hall resistances for integer filling factors in a counterflow (CF) experiment on a density balanced 2D bilayer system. Filling factor dependent equilibration lengths demonstrate enhanced 1D coupling via edge-channels. Due to the narrow barrier the edge-modes of the two 2DEGs are in close proximity allowing for 1D excitonic correlations. Electron drag measurements confirm the observed quantum state selective coupling between the layers.
\end{abstract}
\maketitle


Semiconductor heterostructures formed by Aluminium (Al), Gallium (Ga) and Arsenic (As) have outstanding material properties allowing for high mobility bilayers of 2DEGs in close proximity. The bilayer coupling depends on the ratio between the intralayer and interlayer Coulomb forces, governed by the electron density and the center-to-center quantum well (QW) separation \cite{Murphy1994a}. In addition, the Coulomb gap suppresses the tunneling of electrons between the layers for increasing perpendicular magnetic fields \cite{Eisenstein1992c}. The interaction between adjacent two dimensional charge layers leads to a multitude of phenomena. In zero magnetic field, the electron tunneling between similar 2DEGs shows a conductance resonance at equal densities and zero interlayer bias due to momentum and energy conservation \cite{Eisenstein1992b, Jungwirth1996b,Tsui1994}. In high magnetic fields where the energy levels of the charges are quantized in Landau levels (LLs), a charge condensate state can be observed if the sum of the LL fillings of the adjacent layers equals one. In this case, vanishing Hall and longitudinal resistances in a CF experiment as well as quantized Coulomb drag are signatures of an exciton condensate \cite{Kellogg2004, Kellogg2002}. The latter is made up of filled and empty electron states showing a BCS-like behavior interpreted as condensation of 2D excitions \cite{Eisenstein2014,Spielman2000,Tutuc2004a, Tiemann2009}. Dissipationless flow of the exciton condensate and the Josephson effect between the layers have been observed \cite{Fogler2001}. 

However, the reported research has focused on the condensed state with limited investigations on bilayer samples with separated contacts going beyond its narrow parameter space in terms of electron mobilities and enlarged tunneling couplings. Our double QW sample featuring a 6 nm thin barrier extends the probed parameter range, thereby allowing for a large tunneling coupling of $\Delta_{SAS}\approx 160 \textit{ }\mu V$ while maintaining high electron mobilities. Here, we show quantum state dependent nearly vanishing Hall and longitudinal resistances for integer filling factors in a CF experiment. The coinciding Hall and longitudinal resistive minima in both layers are sustained down to integer filling factors $\nu > 1$. In addition, we present drag resistance measurements which depict the same quantum Hall selective coupling between the layers. 

We establish patterned back gates by photo-lithography and oxygen ion implantation on a MOCVD prepared GaAs wafer \cite{Berl2016a}. Subsequently, the implanted wafer is overgrown by molecular beam epitaxy with an AlGaAs-GaAs heterostructure, featuring two GaAs quantum wells (QW) of 18.7 nm width and an Al$_{0.8}$Ga$_{0.2}$As barrier of 6 nm \cite{Scharnetzky2020,Berl2016a}. The QWs are remotely doped by Silicon, enabling mobilities of $2\cdot10^6\textit{ } cm^2/Vs$ in each QW at a single layer density of $0.92\cdot 10^{11} \textit{ } e/cm^2$.
\begin{figure}
    \centering
    \includegraphics[]{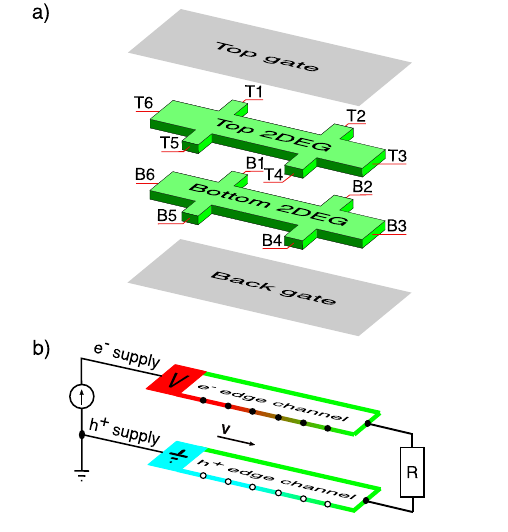}
    \caption{\textbf{a)} Schematic of the bilayered Hall bars. \textbf{b)} Simplified schematic of the Hall bar-shaped edge-modes for a CF experiment.}
    \label{fig: sample_and_edge}
\end{figure}

A Hall bar of 1250 $\mu$m length and 200 $\mu$m width is photo-lithographically processed. We use ion implanted pinch-off back gates and metallic top pinch-off gates to contact the 2DEGs separately \cite{Scharnetzky2020}. Global top and back gates are used to tune the upper and lower 2DEG densities, respectively.
\begin{figure*}
    \centering
    \includegraphics[]{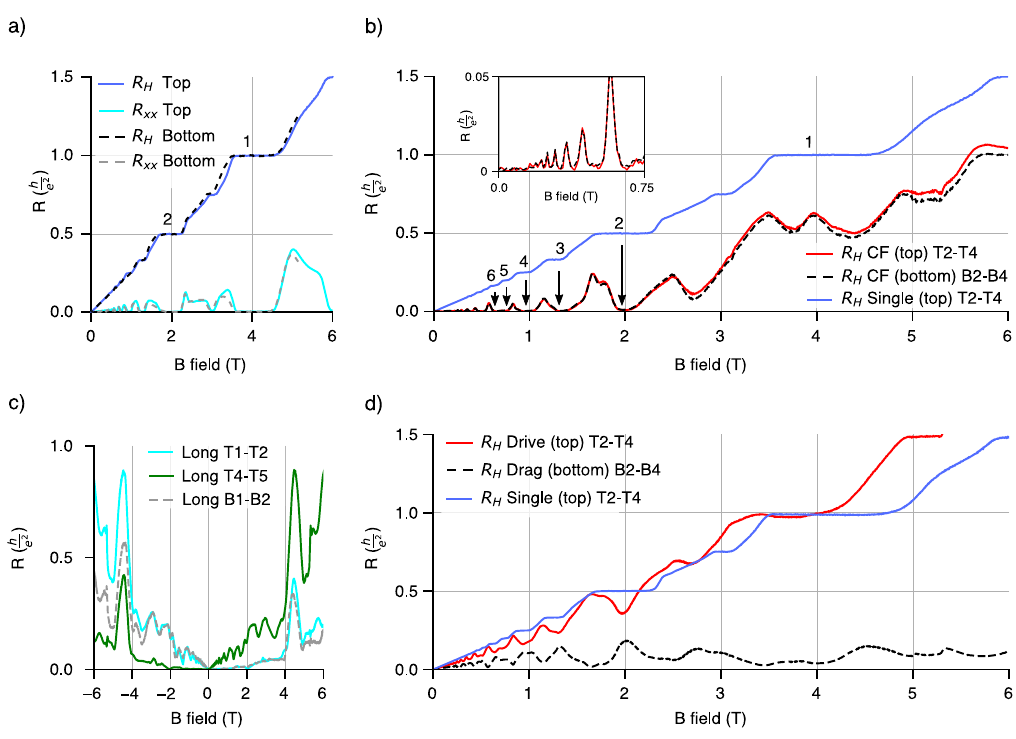}
    \caption{\textbf{a)} Individually recorded Hall and longitudinal resistances for the top and bottom 2DEG. \textbf{b)} The Hall resistances for both layers in the CF experiment nearly vanishes for the same integer fillings. For higher magnetic fields the CF Hall resistance increases. Inset: CF Hall resistances for small fields. \textbf{c)} The longitudinal resistance is asymmetric in magnetic field. The resistances on the same side of the Hall bar are alike. \textbf{d)} The Hall resistance of the current carrying top layer roughly follows the single layer Hall resistance trace. Minima occur at integer filling factors where the drag resistance in the bottom layer features a maxima. The drag resistance does not have a maxima at $\nu = 1$.}
    \label{fig:sample}
\end{figure*}

By applying a sufficient negative back gate voltage the bottom 2DEG can be depleted of its charge carriers. Thereby, the top 2DEG's transport characteristics can be probed as a single layer. The density of the top layer predominately depends on the top gate voltage and is independent of the lower 2DEG density. Analogously, the top 2DEG can be depleted and the bottom QW can be characterized, as seen in \ref{fig:sample} a). Typical quantum Hall resistances featuring plateaus at fractions of the Klitzing constant are obtained \cite{Klitzing1980}, while the longitudinal resistance vanishes at integer fillings. Using this single layer characterization technique, the gate voltages to balance the 2DEG densities can be determined.

By applying Fourier transform analysis to the Shubnikov-de-Haas (SdH) oscillations the energy splitting $\Delta_{SAS}$ between the symmetric and antisymmetric wave functions can be determined \cite{Lo2002}, which for our sample yields a value of $\Delta_{SAS}\approx 160 \textit{ }\mu eV$  (see supplementary fig. 1).

The zero bias tunneling conductivity can be determined by a differential conductance measurement. The zero field conductivity is the largest and decreases strongly with applied perpendicular magnetic field, ranging from ~19 to below 1 mS respectively (see supplementary fig. 2). 

In our CF experiment a current of 1 nA is fed in to the top 2DEG (T6) and is drawn from the bottom 2DEG (B6). A wire connects the far end of the two Hall bars (T3 and B3). The gates are adjusted such that both 2DEGs have the same density of $0.92\cdot10^{11} e/cm^2$.

For a CF experiment with completely uncoupled QWs one would expect the Hall resistivities in the individual layers to coincide with those traditionally observed in a single layer. If, however, the QWs are strongly coupled, meaning that charges can be exchanged readily, then the Hall voltages would compensate each other to zero. The coupling between the two layers and in turn the tunneling conductivity depends on the center-to-center QW distance, the barrier width and on the applied magnetic field. In our sample the tunneling area of 0.25 $mm^2$ results in a tunneling conductivity as high as 19 mS for small perpendicular magnetic fields and lower for higher fields. Hence, for a CF Hall resistance at small fields a strongly reduced Hall resistance due to electron tunneling is expected. For high magnetic fields where the Coulomb gap suppresses tunneling a resistance closer to the single layer resistance has to be anticipated. 

\begin{figure}
    \centering
    \includegraphics[]{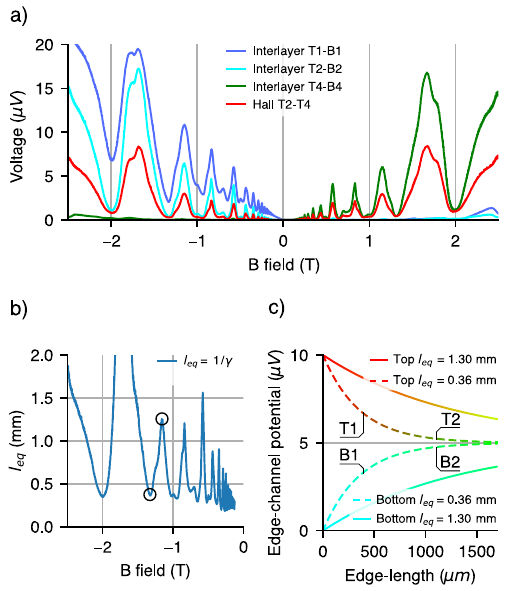}
    \caption{\textbf{a)} Interlayer voltages are strongly asymmetric. For the non-vanishing side, the interlayer voltage is twice the Hall voltage. The interlayer voltage features the same minima as the Hall voltage. \textbf{b)} Calculated equilibration length $l_{eq}$ using the interlayer voltages from one side of the sample are depicted. The equilibration lengths for two fillings $\nu = 3$ and $\nu = 3.5$ are circled. \textbf{c)} Using the circled equilibration length 0.36 mm and 1.3 mm the potentials for the top and the bottom edge-channels are computed.}
    \label{fig: equi_length}
\end{figure}
For our CF Hall resistance measurement we measured the Hall resistance in both layers simultaneously between contacts T2-T4 and B2-B4. In fig. \ref{fig:sample} b) the CF Hall resistances are compared to the single layer Hall resistance with labelled integer filling factors. We record Hall resistances remaining below 100 $\Omega$ for magnetic fields up to 0.2 T. From 0.2 T on until 2.2 T the Hall resistance increases and features quantum state dependent minima smaller than 270 $\Omega$. For filling factors lower than $\nu < 2$ the CF Hall resistance does no longer nearly vanish, but increases with the same slope as the single layer resistance.

Longitudinal resistance and interlayer voltage measurements for the CF experiment both display a pronounced magnetic field sign dependence. The interlayer voltage is composed of either twice the Hall voltage or zero. Whether the voltage is zero or corresponds to twice the Hall voltage depends on the side of the Hall bar, see fig. \ref{fig: equi_length} a). For non-zero potentials the voltage difference follows the same oscillatory maxima and minima as the Hall resistance. Comparing the interlayer voltages on the same side shows that for contacts closer to the current driving (T6) and the ground lead (B6) a larger voltage is measured. Similarly, the longitudinal resistance features a strong magnetic field dependent asymmetry, as seen in fig. \ref{fig:sample} c). Either, the longitudinal resistance increases with magnetic field as 1/B periodic SdH oscillations or is nearly zero depending on the sign of the magnetic field. 

The longitudinal resistance near zero field ($0\pm 0.2\textit{ T}$) is roughly 25 $\Omega$ which is smaller than the corresponding single layer longitudinal resistance. This indicates that a large portion of the current tunnels between the 2DEGs for such small fields. In this magnetic field range the experiment does not resemble a CF experiment. 

We perform an electron drag experiment by driving a\linebreak 1 nA current through the top 2DEG from contact T6 to T3 and measure the Hall resistance in both layers (contacts T2-T4 and B2-B4). In fig. \ref{fig:sample} d) we compare the single layer Hall resistance to the current carrying top layer and the drag resistance in the lower 2DEG. The top layer follows the single layer trace. At integer fillings the resistance drops substantially. For the electron populated but non-driven bottom layer the Hall resistance also increases with magnetic field, with distinctive maxima at integer filling factors. The relative magnitude of the resistance of the bottom layer compared to the top layer indicates the coupling strength between the layers. The locations of the maxima coincide with the minima of the vanishing Hall resistances measured in the CF experiment. 

Condensation of the 2DEGs density of states for a perpendicular magnetic field into LLs can be described by the Büttiker 1D edge-channel picture \cite{Buttiker1988}. When the bulk of the sample is in an insulating gapped state, corresponding to the Fermi level being between two LLs, only the chiral edge-modes carry current. An applied bias offsets the Fermi levels slightly and leads to a net current. In this situation the electron density on one side of the sample is increased and lowered on the other. The density reduction is equivalent to a population with holes.

In the special case of a CF experiment, the electron density increase in the top layer is situated above the augmented hole population in the bottom layer, as illustrated in fig. \ref{fig: sample_and_edge}. This applies for both sides of the Hall bar. As excitonic binding energy of electrons and holes in the 1D is strongly enhanced in contrast to its counterpart for bulk 2D excitons \cite{Bastard1982,Bondarev2011}, we also expect increased tunneling between the 1D edge-channels.

Fig. \ref{fig:sample} b) shows that the Hall resistance nearly vanishes in both layers for an insulating bulk. For the Hall resistance to vanish the potentials across the Hall bars must be equal. This charge balancing cannot take place in the insulating bulk of the sample. In addition, the recorded interlayer voltage is minimal for the same filling factors $\nu$ as the Hall resistance. Therefore, the potentials are nearly the same in both 2DEGs. The Coulomb gap suppresses electron tunneling in the bulk leaving the equilibration of the potentials to the edges. We suggest that the equilibration of the edge-modes at integer filling factors leads to the almost complete vanishing of the Hall resistances observed in the CF experiment.

Equilibration between the edge-modes is established by electron tunneling. A shorter equilibration length relates to an increased coupling between the two 2DEGs. In fig. \ref{fig:sample} d) the drag resistance features distinctive maxima for integer filling factors $\nu \geq 2$ where the current driven top 2DEG shows minima. The increase (decrease) in Hall voltage is caused by an increased (decreased) current in the bottom (top) 2DEG caused by electron tunneling. We suspect that tunneling occurs predominately between edge-channels, where occupied conduction band states are positioned above unoccupied states in close proximity. The electrons in the bulk are localized and there are no unoccupied states. The highest tunneling coupling is achieved at integer filling factors. 

Equilibration between adjacent edge-modes by tunneling has been discussed before \cite{Nicoli2022}. Coupled differential rate equations yield an equation for the   potential difference between the channels 
\begin{equation}
    \label{eq: equilibration}
    \mu_1(x) - \mu_2(x) = e^{-\gamma x}\left[\mu_1(0) - \mu_2(0)\right].
\end{equation}
Where $\gamma$ is the equilibration rate such that $\gamma = p/dx$ is the electron tunneling probability per length. The characteristic equilibration length is then defined by\linebreak $l_{eq} = 1/\gamma$. 

For our bilayer system the left hand side of eq. (\ref{eq: equilibration}) is the measured interlayer voltage at a distance $x$. The right hand side is the initially applied bias times an exponential decay factor depending on the length of copropagation of the edge-channels.
We use eq. (\ref{eq: equilibration}) and the measured interlayer voltages T1-B1 and T2-B2 from fig. \ref{fig: equi_length}  a) to calculate the equilibration length in our CF experiment. The distance between T1 and T2 is\linebreak $\approx 650$ $\mu m$. The equilibration lengths are plotted in\linebreak fig. \ref{fig: equi_length} b). Using the equilibration lengths of 360 $\mu m$ and 1300 $\mu m$ the edge-potentials for an initial 10 $\mu V$ potential difference are calculated with eq. (\ref{eq: equilibration}) and depicted in fig. \ref{fig: equi_length} c). The exponential decay (increase) of the potential along the top (bottom) edge resembles the measured CF resistances and voltages well. The expected interlayer voltage is obtained by choosing a distance on the x-axis and taking the difference between the top and bottom curves. The longitudinal voltage can be constructed by choosing two points on the x-axis and subtracting the computed potentials. Finally, the Hall voltage results by taking the difference of the computed potential to half of the applied voltage. 

From the computed edge-channel potentials of fig. \ref{fig: equi_length} c) the asymmetric behavior of the interlayer voltages and the longitudinal resistances become evident. The edge-modes on one side of the sample have equilibrated to half of the applied potential. Hence, the longitudinal and the interlayer voltage are minimal on that side. With a change in chirality the opposing side of the Hall bar then measures zero potential difference. The nearly vanishing Hall resistance is a consequence of the decreased equilibration length between the edge-modes. 

As described before excitonic correlations between the 1D edge-channels can be responsible for an enhanced coupling at integer filling factors and an almost complete vanishing of the Hall, longitudinal and interlayer voltage. Since for completely filled LLs, where the bulk is insulating, electronic transport exclusively takes place along the sample edges, 1D exciton formation there has the strongest effect on tunneling. 

Strikingly, our picture does not apply to the filling factor $\nu = 1$. The spin polarized filling factor $\nu = 1$ is different. Around $\nu = 1$ skyrmionic ground states consisting of charged spin-textures affect the tunneling between the layers leading to an increase in equilibration length \cite{Sondhi1993,Groshaus2004}. As seen in fig. \ref{fig:sample} b) the Hall resistance does not vanish at filling factor $\nu = 1$. The plateaus at the fraction of the Klitzing constant are present and less pristine. The curve seems to be shifted by the magnetic field corresponding to the filling of $\nu = 2$. The drag resistance in\linebreak fig \ref{fig:sample} d) shows no maxima at filling factor $\nu = 1$, this is compliant with the non-vanishing Hall voltage. The coupling between the two layers remains low for fillings $\nu < 2$. Additionally, the longitudinal resistance partly looses its asymmetry for $\nu < 2$. Therefore, the edge-channels do not equilibrate, which is consistent with a longitudinal resistance measured on both sides of the Hall bar. Tunneling experiments for fillings below $\nu < 2$ must, thus, be described  differently.  

Finally, we do not expect to observe a bulk Bose--Einstein condensate (BEC) with vanishing Hall voltages at filling $\nu = 1/2$ in our counterflow measurement. For our sample the ratio $\Delta_{SAS}/(e^2/(4\pi\epsilon\epsilon_0 l_B))$  is $\approx 12.6e^{-3}$. The condition for a BEC to form in this case is $d/l_B < 2$, which is substantially smaller than our value of 2.65 at 8 T ($\nu = 1/2$).

We widen the observable parameter space by investigating a high mobility sample with a 6 nm barrier and independent contacts to each layer. In our CF experiment we explore the transition between strong and weak couplings of the QWs, where potential equalization takes place through tunneling between the chiral edge-modes of the QWs. We observe nearly vanishing Hall resistances at integer filling factors $\nu > 1$. We route the selectiveness of the equilibration to the enhanced tunneling coupling at filled LLs, which corresponds to a merging of occupied and unoccupied electron states in close proximity at the edge. The increased coupling originates from the excitonic binding between the electrons and holes in the one-dimensional edge-modes. With the currently realized sample design allowing for independent contacts to strongly coupled bilayer systems, investigations on couplings between different filling factors or fractional edge-modes become feasible.

We thank Thomas Ihn and Ady Stern for illuminating discussions. We acknowledge financial support from the Swiss National Science Foundation
(SNSF) and the National Center of Competence in Science "QSIT - Quantum Science and Technology".

%

\end{document}


\title{\label{sec:Supplement}Supplemental material: Strongly interacting 2D electron systems: Evidence for enhanced 1D edge-channel coupling}
\author{C. Marty$^{1,2}$, C. Reichl$^{1,2}$, S. Parolo$^{1,2}$, I. Grandt-Ionita$^{1,2,3}$, J. Scharnetzky$^{1,2}$, W. Dietsche$^{1,2,4}$ and W. Wegscheider$^{1,2}$}
\affiliation{$^1$Laboratory for Solid State Physics, ETH Zürich, Zürich, Switzerland\\$^2$Quantum Center, ETH Zürich, Zürich, Switzerland\\$^3$Institut für Nanostruktur- und Festkörperphysik, Universität Hamburg, Hamburg, Germany\\
$^4$Max-Plank-Institut für Festkörperforschung, Stuttgart, Germany}
\maketitle

\begin{figure}[h!]
    \centering
    \includegraphics[scale = 1]{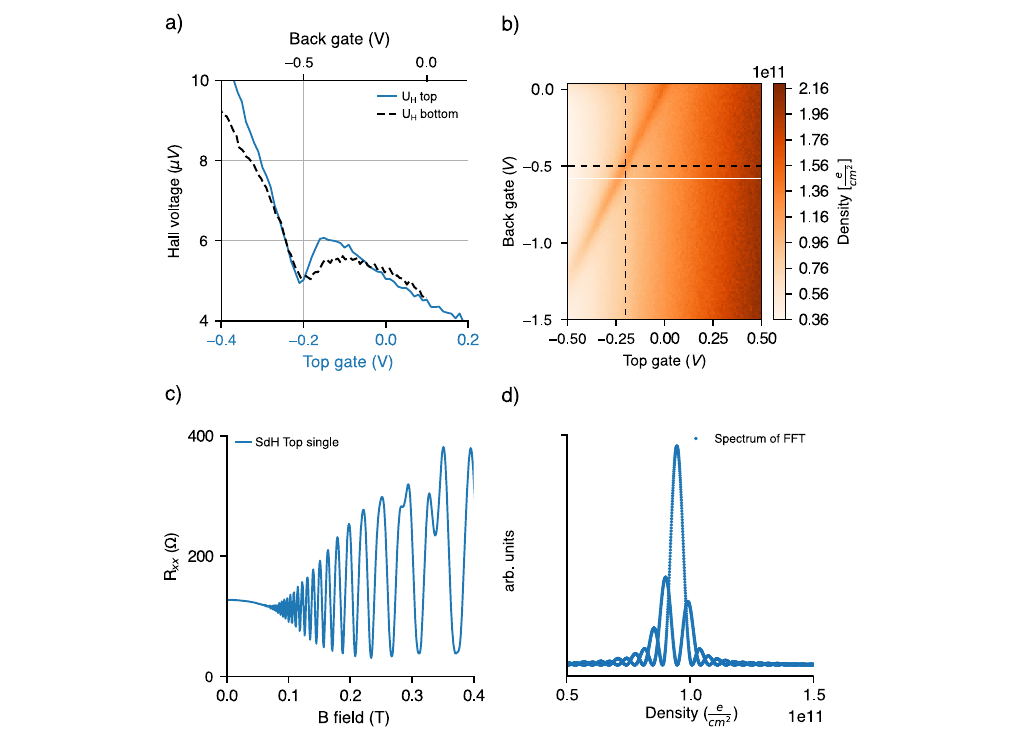}
    \label{fig: supp_first}
    \caption{\textbf{a)} Measured Hall voltage of the top blue (bottom dashed) 2DEG for a top (bottom) gate sweep for a fixed back (top) gate voltage of -0.5 V (-0.2 V) and fixed magnetic field of 0.1 T. A resonant tunneling minima at -0.2 and -0.5 V is apparent at balanced densities. \textbf{b)} Density for the top 2DEG. The density increases linearly with the top gate voltage. At balanced densities resonant tunneling leads to a calculated increased density.\textbf{c)} SdH oscillations of a single top layer. \textbf{d)} FFT of the single layer SdH oscillations from c).}
\end{figure}

\begin{figure*}
    \centering
    \includegraphics[]{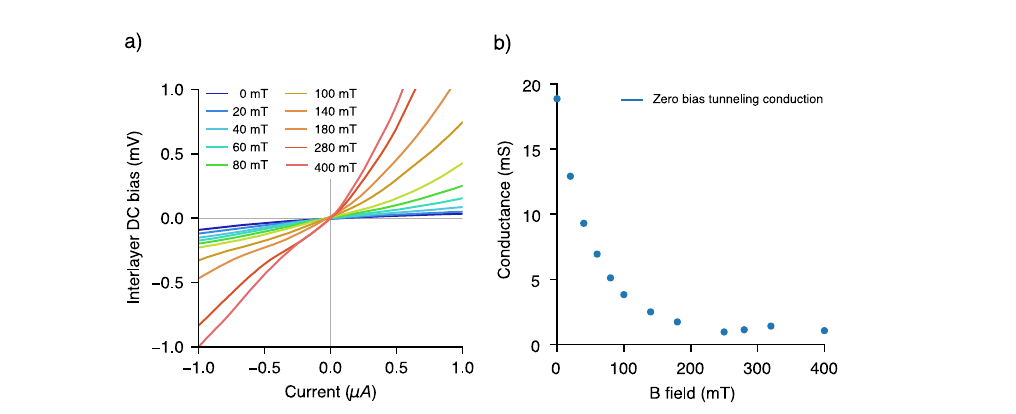}
    \label{fig: supp_second}
    \caption{\textbf{a)} Differential conductance measurements for varying perpendicular magnetic fields. A DC current is forced to tunnel between the two layers which leads to a DC interlayer voltage. \textbf{b)} The dI/dV conductance at zero bias for the fields from a). Tunneling conductance drops with increasing magnetic field.}
\end{figure*}